\documentstyle[twocolumn,prl,aps]{revtex}

\begin{document}
\title{\Large \bf
Exact $Z^2$ scaling of pair production in the high-energy limit 
of heavy-ion collisions} 
\author{B. Segev$^1$ and J. C. Wells$^2$}
\address{
       $^1$Institute for Theoretical Atomic and Molecular Physics, 
       Harvard-Smithsonian Center for Astrophysics,
       60 Garden Street, Cambridge, Massachusetts 02138, USA \\
       $^2$Center for Computational Sciences, Oak Ridge National Laboratory, 
             P.O.\ Box 2008, Oak Ridge, Tennessee 37831-6203, USA}

\maketitle
\begin{abstract}
The two-center Dirac equation for an electron in
the external electromagnetic field of two colliding heavy ions in the 
limit in which the ions are moving at the speed of light is exactly 
solved and nonperturbative amplitudes for free electron-positron 
pair production are obtained.
We find the condition for the
applicability of this solution for large but finite collision energy, and
use it to explain recent experimental results. 
The observed scaling of positron yields as the square of the 
projectile and target charges is a result of an exact cancellation of 
a nonperturbative charge dependence
and holds as well for large coupling. 
Other observables would be sensitive to nonperturbative phases.
\end{abstract}

 {\em PACS number}: 34.50.-s, 25.75.-q, 11.80.-m, 12.20.-m
\vspace{0.1in}

There is a very small number of problems in physics that yield to an
exact solution. Remarkably, 
electromagnetic production of free electron-positron
pairs in the high-energy limit of peripheral, heavy-ion collisions
can be described by a two-center, time-dependent Dirac equation
which can be solved exactly and in closed form\cite{SW98}.
In this Letter, we study this exact solution and discuss its implications 
for recent experiments performed at CERN's SPS \cite{VD}, and 
possible future experiments at new facilities such as BNL's RHIC and 
CERN's LHC. 
(See Refs.\ \cite{greiner,eichler} for relevant reviews of this field.)

Perturbative calculations have been held suspect
at high energies and for the heaviest projectiles,
e.g.\ because the coupling constant is not small ($Z\alpha \sim 
0.6$)\cite{VD,greiner,eichler,BS89a,BaurRB}.
It is therefore surprising that positron yields observed from pair 
production in peripheral collisions of $\rm Pb^{82+}$ ions at $33$-TeV on 
a $\rm Au$ target scale as $Z_T^2Z_P^2$,
and that the observed positron-momentum distributions display
an overall good agreement with the leading-order perturbation-theory 
calculations\cite{VD}.
The exact nonperturbative solution presented here explains these effects
and is also consistent with the observed enhancement of the positron 
yields at small positron momentum. For future experiments,
we indicate what observables would show complete agreement with 
second-order perturbation theory, and what other observables should be 
measured in order to detect nonperturbative effects.

The relativistic scattering problem of an electron in the 
external field of two point-like charges (ions), moving on 
parallel, straight-line trajectories in opposite directions at 
speeds which approach the speed of light, and at an 
impact parameter $2 \vec{b}$,
reduces in the high-energy limit to
\begin{eqnarray}
i\frac{\partial}{\partial t} |\Psi(\vec{r},t)\rangle &=&
\left[ \hat{H}_0 + \hat{W}_A(t) + \hat{W}_B(t) \right] 
|\Psi(\vec{r},t)\rangle , \nonumber \\
\hat{H}_0 &\equiv& -i \check{\alpha}\cdot\vec{\nabla} + 
\check{\gamma^0} , \nonumber \\
\lim_{\gamma \rightarrow \infty } \hat{W}_A &=&
 ({\rm I}_4-\check{\alpha}_z) 
Z_A \alpha \delta (t-z) \ln \left[ 
\frac{(\vec{r}_{\perp}-\vec{b})^2}{b^2} \right] 
,  \nonumber \\
\lim_{\gamma \rightarrow \infty }
     \hat{W}_B &=& ({\rm I}_4+\check{\alpha}_z) 
Z_B \alpha \delta (t+z) \ln \left[ 
\frac{(\vec{r}_{\perp}+\vec{b})^2}{b^2} \right]\;,
\label{dirac}
\end{eqnarray}
where $\gamma \equiv 1/\sqrt{1-\beta^2}$, and $\beta \equiv v/c$ is
the speed of the charges, $Z_A$ and $Z_B$. 
($\gamma$ in the collider frame is 
related to $\gamma_T$ in the target frame by 
$\gamma_T=2\gamma^2-1$.)
Equation (\ref{dirac}) is written in 
the collider frame, with natural units ($c=1$, $m_e=1$, and $\hbar=1$), 
$\alpha$ is the fine-structure constant, and
$\check{\alpha}$ and $\check{\gamma}^{\mu}$ are Dirac matrices 
in the Dirac representation.
The $\delta$-function form of the interaction is a
high-energy limit of the exact interaction in 
a {\it short-range} representation for the electron's
Dirac-spinor, chosen to remove the 
interaction at asymptotic times\cite{Ja92,Baltz97,WS98}.

For Eq.\ (\ref{dirac}) to apply,
one assumes first that 
the ions are sufficiently energetic and massive so that the 
deviation from straight-line trajectories can be neglected\cite{eichler}.
Second, one assumes peripheral collisions without nuclear interaction.
Purely electromagnetic events can be distinguished experimentally from
nuclear events by observing the full-energy projectile ion
after the collision in coincidence with the produced
electrons or positrons \cite{VD}. 
One also assumes the ions are moving at the speed of light. 
Indeed, for the recent experiments at CERN \cite{VD}, 
$\beta \approx 0.99$,
and for future experiments possible at RHIC, $\beta \approx 0.9999$.
Finally, to obtain the $\delta$-function form of the 
interaction, one assumes that 
$\gamma\gg |\vec{r}_{\perp}\pm \vec{b}|$, and $ 2b$.

Equation
(\ref{dirac}) displays a unique electromagnetic interaction. 
The ion with 
charge $Z_A$ is moving to the right at the speed of light. 
Its electromagnetic potential
in the representation chosen here is 
Lorentz contracted to the plane transverse to its trajectory, (the 
light front $z=t$), hence the $\delta(z-t)$ functional dependence. 
Likewise, the ion with charge 
$Z_B$ carries with it at the other light front, ($z=-t$), a plane of 
singular interaction moving to the left at the speed of light. 
Anywhere 
but on the light fronts, Eq.\ (\ref{dirac})  reduces to the free Dirac 
equation.
The Dirac plane waves $\{|\chi_p(\vec{r},t)\rangle
= \exp (-i E_p t)
\exp(i \vec{r}\cdot\vec{p}) |u_p \rangle \}$ which satisfy the free 
Dirac equation are each characterized by the quantum numbers
$p\equiv\{\vec{p},\lambda_p,s_p\}$;
the momentum $\vec{p}$, the sign of the energy 
$E_p= (-1)^{\lambda_p} \sqrt{p^2+1}$,
and the spin $s_p=\pm$.
Explicit forms for the four four-spinors $|u_p \rangle $ are given in 
Ref.\ \cite{eichler}, and $p_\pm\equiv p_z\pm E_p$.

The scattering problem for the electron is defined by Eq.\ (\ref{dirac})
and by plane-wave asymptotic initial and final states.
One solves for the transition amplitude $A^{(j)}_k$,
   \begin{eqnarray}
&&\lim_{t\rightarrow -\infty}|\psi^{(j)}(\vec{r},t)\rangle
=|\chi_j(\vec{r},t)\rangle
\nonumber , \\
&&\lim_{t\rightarrow +\infty}|\psi^{(j)}(\vec{r},t)\rangle
=\sum_{k} A^{(j)}_k |\chi_k(\vec{r},t)\rangle ,
\label{amplitude}
\end{eqnarray}
where $\sum_{k}$ stands for integration over $\vec{k}$ and 
summation over $\lambda_k$ and $s_k$. 
We have obtained an exact, closed-form integral representation for the 
scattering amplitude $A^{(j)}_k$ in the following manner\cite{SW98}. 
First we have observed that,
as the ions are approaching from infinity, no change
occurs in the region of space between 
them ($|z|<|t|$) until $t=0$ when the 
two singular interaction planes collide. 
However, as each ion sweeps through space, it interacts
with the single plane wave it encounters, resulting in
a superposition of plane waves after it passes.
Each $\delta$-function interaction induces a 
phase-shift discontinuity in the wavefunction
across each light front\cite{Ja92,Baltz97,SW98}. 
A phase shift induced on a plane wave by the passage of
a single ion is not sufficient to produce pairs,
but as the two phase-shift planes collide at $t=0$,
they interfere, and pairs are produced as a result.
After the collision ($t>0$), as the ions move apart,
the solution in the space between them ($|z| < t$)
is a new superposition of plane waves which is
determined by the nontrivial boundary condition at the light fronts.
We have calculated the transition amplitudes $A^{(j)}_k$ by
integrating the flux of the conserved transition four-current
which flows into this region across the light fronts.
Two terms contribute to the amplitude corresponding to the two
time orderings of the interaction of the electronic
wavefunction with the two ions.

The transition amplitudes, $A^{(j)}_k$, are represented in terms of the 
transverse-momentum transfer distribution induced by a single ion, 
$Q^{\vec{b}}_Z(\vec{\kappa})$, which
contains all the dynamics of the 
ion-electron interaction. 
When $\lambda_k=0$ and $\lambda_j=1$, $A^{(j)}_k$ is an amplitude for a 
transition from the negative continuum to the positive continuum, i.e.\ 
an amplitude for pair production. We have found\cite{SW98},
   \begin{eqnarray}
A^{(j)}_k &=& \frac{i}{\pi}
\int d \vec{p}_{\perp}
\left\{ 
\frac{\sigma^j_{k}(\vec{p}_{\perp})
\ Q^{\vec{b}}_{Z_B} (\vec{k}_{\perp}-\vec{p}_{\perp})
\ Q^{\vec{b}}_{Z_A} (\vec{j}_{\perp}-\vec{p}_{\perp})
}{p^2_{\perp}+1-j_-k_+}
      \right. \nonumber \\  && \left. -
\frac{\sigma^{k \dagger}_{j}(\vec{p}_{\perp})
\ Q^{\vec{b}}_{Z_A} (\vec{p}_{\perp}-\vec{k}_{\perp})
\ Q^{\vec{b}}_{Z_B} (\vec{p}_{\perp}-\vec{j}_{\perp})
}{p^2_{\perp}+1-j_+k_-}
\right\}.  \label{amp}
\end{eqnarray}
The spinor part is
$\sigma^j_{k}(\vec{p}_{\perp})\equiv (2\pi)^3
\langle u_k| (I_4-\check{\alpha}_z)
(\check{\alpha}\cdot
\vec{p}_{\perp}+\check{\gamma}^0) (I_4+\check{\alpha}_z) 
| u_j\rangle$ 
and the momentum-transfer distribution is given by 
the Fourier transform of the phase shift at the light front\cite{SW98},
   \begin{eqnarray}
Q^{\vec{b}}_{Z}(\vec{\kappa})&\equiv&
\frac{1}{(2\pi)^2}\int d \vec{r}_{\perp}
\ e^{i \vec{r}_{\perp}\cdot\vec{\kappa}}
\left[\frac{(\vec{r}_{\perp}-\vec{b})^2}{b^2}\right]^{- i \alpha Z}
\nonumber \\
&=& \frac{1}{2\pi}
\frac{\exp (i \vec{\kappa}\cdot\vec{b})}{\kappa^2
(b \kappa)^{- i 2\alpha Z} }
\int_{\xi>0} d\xi J_0(\xi) \xi^{1- i 2\alpha Z} , \label{Q}
\end{eqnarray}
where $Z$ is the charge of the corresponding ion and $\vec{\kappa}$ is the 
transverse-momentum transfer.
The integral over $\xi\equiv\kappa |\vec{r}_{\perp}-\vec{b}|$ 
in Eq.\ (\ref{Q}) should be regularized so as to avoid
unphysical contributions from large, transverse distances,
i.e. from $\xi > \kappa \gamma$.
We studied several 
regularization schemes (for $\vec{\kappa}\neq 0$) 
which all gave\cite{watsonbook}
   \begin{eqnarray}
&&Q^{\vec{b}}_{Z}(\vec{\kappa})\rightarrow
\frac{-i\alpha Z}{\pi}
\frac{\exp (i \vec{\kappa}\cdot\vec{b})}{\kappa^2}
\left[
\frac{\Gamma(-i \alpha Z)}{\Gamma(+i \alpha Z)} 
\left(\frac{b \kappa}{2}\right)^{+ i 2\alpha Z} 
\right].
\label{mtd}
\end{eqnarray}

The exact amplitudes in the infinite $\gamma$ limit are obtained by
substituting the result of Eq.\ (\ref{mtd}) in Eq.\ (\ref{amp}),
   \begin{eqnarray}
\lim_{\gamma\rightarrow\infty} A^{(j)}_k &=& 
\left[
\left(\frac{b}{2}\right)^{+ i 2\alpha (Z_A+Z_B)} 
\frac{\Gamma(-i \alpha Z_A)}{\Gamma(+i \alpha Z_A)} 
\frac{\Gamma(-i \alpha Z_B)}{\Gamma(+i \alpha Z_B)} 
\right]
        \nonumber \\ && \times 
 \frac{i}{\pi^3} \ \alpha^2 Z_A Z_B
\int d \vec{p}_{\perp}
 {(\vec{p}_{\perp}-\vec{k}_{\perp})^{-2}
\ (\vec{p}_{\perp}-\vec{j}_{\perp})^{-2}}
        \nonumber  \\ &&  \times 
\left\{ 
\frac{\sigma^j_{k}(\vec{p}_{\perp})}{p^2_{\perp}+1-j_-k_+}
e^{{i\vec{b}\cdot(\vec{j}_{\perp}+\vec{k}_{\perp}-2\vec{p}_{\perp})}}
        \nonumber  \right. \\ &&  \left. \times 
\left[
|\vec{p}_{\perp}-\vec{k}_{\perp}|^{i 2\alpha Z_A}
|\vec{p}_{\perp}-\vec{j}_{\perp}|^{i 2\alpha Z_B}
\right]
        \nonumber \right. \\ && \left. - \ 
\frac{\sigma^{k \dagger}_{j}(\vec{p}_{\perp})}{p^2_{\perp}+1-j_+k_-}
e^{{-i\vec{b}\cdot(\vec{j}_{\perp}+\vec{k}_{\perp}-2\vec{p}_{\perp})}}
        \nonumber  \right. \\ &&  \left. \times 
\left[
|\vec{p}_{\perp}-\vec{k}_{\perp}|^{i 2\alpha Z_B}
|\vec{p}_{\perp}-\vec{j}_{\perp}|^{i 2\alpha Z_A}
\right] \right\} \; .
\label{amp2}
\end{eqnarray}
The branch-point singularities for the intermediate momentum 
$\vec{p}_{\perp}=\vec{k}_{\perp}$ or $\vec{j}_{\perp}$ are an artifact of 
using Eq.\ (\ref{mtd}) for $\vec{\kappa}=0$.
(The integral over $\vec{p}_{\perp}$ in Eq.\ (\ref{amp}) has no 
singularities.)
We continue the analysis  assuming an 
appropriate regularization at these points.

Equation (\ref{amp2}) 
is nonperturbative, and already includes the 
interaction to all orders in $\alpha Z$ for any value $\alpha Z$ may have.
Yet, its form is very similar to the high-energy limit of results obtained 
from the two-photon exchange diagrams of second-order perturbation 
theory\cite{BS89a}, (which in the following we simply call the 
perturbative result).
For high-energy collisions, there is therefore no reason 
to calculate higher order diagrams.
As $\gamma\rightarrow\infty$, the only corrections to second-order 
perturbation theory calculations for
free-pair production, including both higher orders and nonperturbative
effects, are the {\it phases} in the square brackets. 
For small values of $\alpha Z$,
these phases tend to 1 and the perturbative limit is reproduced\cite{SW98}. 
What are the observable nonperturbative effects for finite charges?
The phase in the square brackets outside the integral over 
$\vec{p}_{\perp}$ has no physical implications, 
but the phases in the integrands
may substantially alter the physical predictions. We 
find, for example, that the high-energy limit for
${|A^{(j)}_{k}|}^2$ differs, in 
general, from the perturbative result. On the other hand, 
using Eq.\ (\ref{amp2}) to calculate the integrated
observable $\int d \vec{(2b)} {|A^{(j)}_{k}|}^2$, 
we get
   \begin{eqnarray}
&& 
\frac{4}{\pi^4} \ \alpha^4 Z_A^2 Z_B^2
\int d \vec{p}_{\perp} \
(\vec{p}_{\perp}-\vec{k}_{\perp})^{-4}
\ (\vec{p}_{\perp}-\vec{j}_{\perp})^{-4}  
\label{prob} \\
&& \times \left\{ 
\frac{\left|\sigma^j_{k}(\vec{p}_{\perp})\right|^2}{(p^2_{\perp}+1-j_-k_+)^2}
+
\frac{\left|\sigma^j_{k}(\vec{p}_{\perp})\right|^2}{(p^2_{\perp}+1-j_+k_-)^2}
\right. \nonumber \\
&& \left. - \ 2 Re 
\frac{\sigma^j_{k}(\vec{p}_{\perp})
\sigma^{k}_{j}(\vec{j}_{\perp}+\vec{k}_{\perp}-\vec{p}_{\perp})}
{(p^2_{\perp}+1-j_-k_+)
((\vec{j}_{\perp}+\vec{k}_{\perp}-\vec{p}_{\perp})^2+1-j_+k_-)}
\right\} \; ,
\nonumber
\end{eqnarray}
which is identical to the perturbative result.
The integration over the impact parameter  
results here in cancellation of the nonperturbative phases, as $\int d 
\vec{(2b)} \exp(i\vec{2b}\cdot\vec{p}')= 
(2\pi)^2 \delta(\vec{p}')$.
We conclude that while some observables are sensitive to the 
nonperturbative phases, other observables are not, 
e.g.\
because these 
phases are averaged to one by an integration.
In these cases, observed 
results would agree with the second-order perturbation theory 
calculations, regardless of the size of $\alpha Z$.

Would our results apply in an actual experiment, where $\gamma$ is 
finite? Equation (\ref{dirac}) is {\it incorrect} 
for large ${r}_{\perp}$ (or large $b$) 
where it describes an interaction which continually
increases in strength.
Implicit in using Eq.\ (\ref{dirac}) for large, finite $\gamma$ 
is a nontrivial assumption that
large, transverse distances do
not contribute to pair production.
In the recent experiments at CERN \cite{VD},
$\gamma\approx 10$,
while in possible future experiments at RHIC and LHC, 
$\gamma\approx 100$ and $\gamma\approx 3000$, respectively.
For these values of $\gamma$,
Eq.\ (\ref{dirac}) is 
respectively limited to pairs produced at transverse distances 
much smaller than 10, 
100, and 3000 Compton wavelengths away from the ions.
This restriction is consistent with an experimental observation 
according to which the average length scale for pair production in 
relativistic heavy-ion collisions is one Compton 
wavelength\cite{VD}. 
Yet, being concerned here not with averages but with complete 
distributions, we can not exclude the possibility of some pairs being 
produced at large, transverse distances from the highly charged ions, 
as long as the transverse-momentum transferred is 
sufficiently small.
For Eq.\ (\ref{mtd}) to be meaningful, 
the regulated integral of Eq.\ (\ref{Q}) must converge to the expression 
of Eq.\ (\ref{mtd}) for $\xi$ such that 
$|\vec{r}_{\perp}-\vec{b}|\ll\gamma$.
The case of small coupling was previously studied\cite{SW98}. 
The case of 
large $\alpha Z$ can be considered by the method of stationary phase. 
Expansion 
of Eq.\ (\ref{Q})
around the stationary point $\vec{r}_{\perp}-\vec{b}= 
{2\alpha Z}\vec{\kappa}/{\kappa^2}$ confirms Eq.\ 
(\ref{mtd}) for this case. 
The procedure is consistent if
the stationary point is located at small distances from the ion, i.e.\
if and only if 
   \begin{eqnarray}
|\vec{\kappa}|\gg \frac{2\alpha Z}{\gamma} \; .
\label{condition}
\end{eqnarray}
It is interesting to find that Eq.\ (\ref{condition}) is trivially 
satisfied in two very different limits: in the perturbative limit of 
$\alpha Z \rightarrow 0$ and in the high-energy limit of 
$\gamma\rightarrow \infty$. 

Thus, the results which we have first obtained for infinite $\gamma$, 
apply for finite $\gamma$ as well. The only restriction is of Eq.\ 
(\ref{condition}), i.e.\ that the transverse-momentum 
transfer is not too small. 
For pair production, it is 
a sufficient condition
to assume that either the initial or final
(i.e.\ positron or electron)
transverse-momenta
are much larger than $2\alpha Z/\gamma$ where $Z$ is the largest 
{\it free} charge involved in the collision.
The argument goes as follows. There are three two-dimensional 
integration variables in Eq.\ (\ref{amp}). We first integrate over 
$\vec{p}_{\perp}$ to obtain simple combinations of the
Bessel functions of the third kind, $K_0$ and $K_1$. 
We then use the 
condition
that {\it one} of the two transverse 
momenta, $\vec{j}_{\perp}$ {\it or} $\vec{k}_{\perp}$, is much larger than 
$2\alpha Z/\gamma$ to apply a stationary phase calculation to one of the 
coordinate integrations. If, on the other hand, one of the charges is 
screened (a target charge, for example) the integral with it converges 
and there is no need to restrict the momentum conjugate to it.
The last integral, over the other 
coordinate-integration variable, converges due to the  Bessel functions 
which drop exponentially for large values of their arguments. 
Having thus proved that contributions for the 6-fold integral of 
Eq.\ (\ref{amp}) from large, transversal coordinates can be neglected, 
we can make the substitution of  Eq.\ (\ref{mtd}) and 
obtain Eq.\ (\ref{amp2}). We 
remark that the convergence of the $\vec{p}_{\perp}$ integration to the 
Bessel functions  
occurs only
for pair-production amplitudes for which 
$1-j_{\pm}k_{\mp}>0$, and is directly related to 
the mass gap between the two continua. 
It should be reconsidered for transitions within the same 
continuum. 

We now consider the application
of our results to the discussion of recent,
pioneering experiments on pair production performed at CERN's SPS
\cite{VD}.
These experiments measured momentum spectra of positrons emitted
from pair production in peripheral collisions of $33$-TeV $\rm Pb^{82+}$
ions ($\gamma_T = 168$) and $6.4$ TeV $\rm S^{16+}$ ions ($\gamma_T=212$)
with various targets (i.e.\ ($\rm CH_2$)$\rm _x$, Al, Pd, and Au).
The charge dependence of the positron yield was reported with
excellent precision.
The target-charge dependence for the sulfur projectile is
$\rm Z_T^{1.99\pm 0.02}$, and for the lead projectile is
$\rm Z_T^{2.03\pm 0.03}$; both within $\sim1\%$ agreement with
the prediction of perturbation theory.
The projectile-charge dependence was observed to be
$\rm Z_P^{2.0\pm 0.1}$, also 
in very good agreement with perturbation theory.
The positron momentum distributions for sulfur
and lead projectiles are compared by scaling each spectrum by $Z_P^2$,
and by scaling the sulfur data from $\gamma_T = 212$ to $\gamma_T =168$
as $\ln^3(\gamma_T)$, as predicted by perturbation theory.
The scaled distributions are observed to be approximately the same, and
to agree reasonable well with two-photon perturbation theory (see
discussion in \cite{VD}),
except for enhancements 
for the lead projectile
at very low ($<2$ MeV/c) and high (between $8$ MeV/c
and $12$ MeV/c) momentum.
The authors of Ref.\ \cite{VD} note that the variation of the
scaled momentum distribution
with the projectile charge,
and not the target charge, is unexplained.

The observed $Z_T^2 Z_P^2$ charge dependence of the single-positron
yields, even for very large charges, is consistent with
the charge dependence we have obtained for the nonperturbative, 
high-energy limit (see Eq.\ (\ref{prob})).
It agrees with perturbation theory but 
is not a perturbative effect.
Nonperturbative phases in the exact amplitudes make them different from 
second-order perturbation theory results, but these phases cancel for 
calculations of total cross sections. 
Our theoretical prediction of a
$Z^2$ dependence of the total cross section in the high-energy
limit implies that multiple-pair production in very high-energy
collisions cannot be inferred from a measurement of the charge dependence
of the total positron yield\cite{VD}.
We suggest that the two regions of excess cross
section observed in the experiment have a common origin:
an enhancement over perturbation theory  for small values
of the transverse-momentum transfer, for which Eq.\ (\ref{Q}) diverges.
We found that agreement with the perturbative
result  is restricted by Eq.\ (\ref{condition}). 
Assuming that in the collider frame pairs are produced isotropically,
this restriction, formulated for the transverse momentum,
may translate to a restriction on the total positron momentum:
$j_\perp \sim j_z \sim j >> 2 Z_P \alpha / \gamma$. Taking $>>$ to be a 
factor of $10$, we then predict, in very good agreement with the 
observed scaled spectra, that the perturbative result 
for the positron momentum distributions is valid 
for $j > 0.4$ MeV/c for the
sulfur data, and $ j > 2 $ MeV/c for the lead data.
The excess cross section observed at high momentum for the lead
projectile is consistent with an enhancement in the cross section
at low transverse momentum in the collider frame after a
relativistic transformation to the target frame is applied.
The absence of an observed target-charge dependence for the scaled 
distributions, is most likely attributable to screening by the atomic 
electrons\cite{screening}. 

In conclusion, we have shown that the exact, nonperturbative solution and
the two-photon exchange diagrams of 
second-order perturbation theory give exactly the same results for 
free-pair production yields integrated over the impact parameter, as 
long as the transverse-momenta transferred
from the ions to the electron
are larger than $2\alpha Z/\gamma$. 
The leading-order perturbative calculations {\it for this observable}
are therefore exact not only at 
the perturbative limit of $\alpha Z\ll 1$ but also in the high-energy 
limit of $\gamma\gg1$. This explains recent 
experimental 
results according to which production rates scale as $Z_P^2 Z_T^2$,
even for large charges.
New nonperturbative effects could be detected by 
measuring observables different from the integrated, inclusive production 
rate that was measured in these experiments. 
The exact amplitudes of Eq.\ (\ref{amp2}) include nonperturbative phases
which may have an observable effect, e.g.\ if one does not integrate over 
the impact parameter $\vec{2b}$.
We expect these phases to strongly influence the theoretical 
predictions for correlations and multiple-pair production.
Several issues deserve further study. These include pair production at 
large, transverse distances from the ions and bound-free production for 
which one should obtain the solution {\it on} the light fronts themselves. 
For other calculations, Eq.\ (\ref{amp2}) as well as the physical picture 
that led to it, are likely to become useful theoretical 
tools\cite{newpreprints}.

\vspace{-0.2in}
\section*{Acknowledgments}
\vspace{-0.1in}

This work was partially supported by the  National Science Foundation
through a grant to Harvard University and the Smithsonian 
Astrophysical Observatory, and by Oak Ridge National Laboratory 
under contract DE-AC05-96OR22464 
with the U.S. Department of Energy.

\vspace{-0.2in}


\begin{thebibliography}{99}
\bibitem{screening}
               Screening begins at the K-shell radius 
               ($a_K$) and is assumed complete at the Thomas-Fermi 
               radius ($a_{TF}$), (see also \cite{So98}).
               As either $a_K$ or $a_{TF}$ may be larger than $\gamma$,
               more work is required for detailed comparison.
               However, since the divergence in
               Eq.\ (\ref{Q}) is slow (logarithmic)
               and the charge dependence of $a_{TF} \sim 1/(Z_T^{1/3}\alpha)$
               is weak, the target-charge dependence of the 
               scaled momentum distributions for fixed-target experiments
               will be very weak.
\end{thebibliography}
\end{document}